%% file: ms.tex
\documentclass[conference,a4paper]{IEEEtran}
\usepackage{cite}
\ifCLASSINFOpdf
  \usepackage[pdftex]{graphicx}
  \usepackage{xcolor}  
\else
\fi

\makeatletter
\def\ps@IEEEtitlepagestyle{%
	\def\@oddfoot{\mycopyrightnotice}%
	\def\@evenfoot{}%
}
\def\mycopyrightnotice{%
	{\footnotesize 20.5.2019}
	\gdef\mycopyrightnotice{}
}
\begin{document}
	
\newcommand{\Quartierstrom}{\textsl{Quartierstrom }}
\newcommand{\Walenstadt}{\textsl{Walenstadt }}

\title{Privacy-Preserving P2P Energy Market on the Blockchain}


\author{\IEEEauthorblockN{Alain Brenzikofer}
\IEEEauthorblockA{Supercomputing Systems AG\\
	alain.brenzikofer@scs.ch}
\and
\IEEEauthorblockN{Noa Melchior}
\IEEEauthorblockA{Supercomputing Systems AG}
}

\maketitle

\begin{abstract}
\Quartierstrom creates a peer-to-peer marketplace for locally generated solar power. The marketplace is implemented as a smart contract on a permissioned blockchain governed by all prosumers. Two privacy-by-design concepts are presented which guarantee that the user’s individual load profile is not leaked to any third party despite using a blockchain. The first approach leverages UTXO based coin mixing protocols in combination with an account-based on-chain smart contract. The second approach relies on an off-chain smart contract running in trusted execution environments.
\end{abstract}

\section{Introduction}

\input{introduction.tex}

\section{Transparent-Bid Auction with Shielded Bid Originators}\label{sec:shieldedtxauction}

\input{shielded-auction.tex}

\section{Off-Chain Auction with Trusted Execution Environments}\label{sec:teeauction}

\input{tee-auction.tex}

\section{Related Work}

\input{related-work.tex}

\section{Conclusion}
We have introduced the privacy challenges when implementing a decentralized auction for electrical energy on a blockchain. Two concepts have been presented that provide confidentiality of personal data while still delivering transparency of the auction process. The first approach breaks the linkability among subsequent orders by means of either zero-knowledge proofs or ring signatures. The second approach leverages trusted execution environments to guarantee both confidentiality and integrity of the auctioneer. Both approaches can be applied to centralized and decentralized systems alike, enhancing privacy with or without leveraging blockchain.

\section*{Acknowledgments}
The \Quartierstrom project is co-funded by the Swiss Federal Office of Energy (SFOE). The consortium consists of Wasser- und Elektrizitätswerke Walenstadt WEW, bits to energy lab ETHZ, Bosch IoT Lab HSG, Hochschule Luzern, Supercomputing Systems AG, Cleantech 21, Sprachwerk, Planar, SwiBi, BKW.

\input{bibliography.tex}
\end{document}

%% file: introduction.tex
Renewable electrical energy is increasingly produced locally in a decentralized fashion. Self-consumption of this produced energy is incentivized in many legislations in order to improve the profitability of renewables. 
However, most prosumers currently have no possibility to influence the level of remuneration for the energy they sell nor are they allowed to sell directly to their neighbors. 
The \Quartierstrom project \cite{quartierstrom} implements a transactive energy system that manages the exchange and remuneration of electricity between consumers, prosumers and the utility in the absence of intermediaries. 
Blockchain \cite{nakamoto08} technology was chosen to implement the decentralized peer-to-peer market that does not rely on a trusted third party (TTP). Both prosumers and consumers can indicate a price at which they are willing to sell or buy locally produced solar energy without the intermediation of a utility or any other TTP. 
Utilities still enjoy a lot of trust in most european countries \cite{eusurvey}. However, this trust mainly regards their integrity, not their competence in information security matters nor the absence of their curiousness regarding personal data. 

Blockchain solutions have the potential to guarantee transparency and integrity of the process along with confidentiality and information security. Moreover, the absence of any single point of failure improves resiliency. 
Finally, blockchain technology is a natural choice to endorse the bottom-up community spirit, attractive to many customers engaging in decentralized energy production.

\subsection{Decentralized Energy Market}
\Quartierstrom implements a double auction with discriminatory pricing as its market mechanism. For both, consumers and prosumers, the smart meters transmit bids containing the price limit determined by the individual household and the electricity demand or supply measured by the meter. 
An order book collects all bids during discrete intervals of 15 minutes and sorts them by price: Sell bids with a lower sell price are prioritized, and buy bids with a higher price, respectively. 
Discriminatory pricing means that for each trade, the price is derived as the mean between the buyer's and seller's price. 

Befitting the idea of decentralization, the auction is implemented as a smart contract on the blockchain. \Quartierstrom is built upon a Tendermint BFT consensus \cite{tendermint} with nodes running on embedded devices \cite{meeuw} at the metering points in the grid.

\subsection{Privacy Challenge}
The aggregated power consumption of each individual household is sampled every 15 minutes and placed on the market as a bid by means of a blockchain transaction. 
Linking bids for a particular participant reveals his usage profile, which is to be considered personal data. 
A Nakamoto blockchain \cite{nakamoto08} guarantees pseudo anonymity by representing each demand and production smart meter as a public key (the public key can be thought of as an account number in the traditional banking sense). 
No association between the public key and the household's address or occupant's identity is published on the blockchain at any point in time. 
However, third parties may be able to gain insights about consumer behavior, household characteristics or occupancy patterns from market orders. Because of this linkability risk, the European Blockchain Observatory considers public keys personal data under GDPR \cite{gdpr}. 

\subsection{Goal}
\Quartierstrom therefore aims at breaking the linkability among subsequent market orders. 
The \Quartierstrom marketplace features a public order book, therefore price and amount of all orders are public. 
This transparency is desired so every observer can verify the market's integrity. However, the identity of a bid's originator does not need to be public as long as the market can enforce settlement of successfully cleared bids. Leveraging blockchain features, a cryptocurrency can be used to supply the necessary funds along with a bid using an atomic transaction. Such a cryptocurrency needs to feature private transactions to avoid linkability of bids.
As the group of users on a market is bound by the physical dimensions of the respective grid, k-anonymity \cite{samarati2001} can be achieved at best, k being the number of participants per grid region. 
This paper presents two fundamentally different approaches that were evaluated as candidates for privacy enhancements of the \Quartierstrom system. Section \ref{sec:shieldedtxauction} approaches the problem with cryptography, while section \ref{sec:teeauction} leverages trusted execution environments.

%% file: shielded-auction.tex
\label{sec:shieldedauction}
\Quartierstrom implements a double auction by means of a smart contract. 
The business logic behind this marketplace is not further explored here. 
Instead, the focus is put on user privacy in a transparent market on blockchains in general.

Orders contain a price in [cts/kWh] and the amount of energy in [kWh] consumed or produced during the last 15 minutes. As the order is sent to the market as a blockchain transaction, the sender has to sign the order with her private key. Therefore, her public key is tied to the order, which leads to unwanted information leakage as subsequent orders can be linked to the same public key, thereby revealing consumptional behavior.  

On a blockchain, the bidder can directly supply the maximum value of her order together with the bid, in the form of cryptocurrency. If her order is cleared at a lower price, the change can be returned to the sender address. Supplying the necessary value along with the order is the first step to unlink orders because it makes every order binding by itself and the market contract can settle all trades directly without knowing the identity or an account number of the originator. 
However, even if one may now use a different address for every order, one still has to fund these addresses, thereby linking them. As shown in \cite{reid12}, tracing blockchain transactions can easily reveal a sender's identity. 

We propose to use shielded addresses to break linkability between orders. Shielded addresses are shadow identities unlinkable to the main addresses, which can be used to transfer tokens without revealing the originators, the transaction amount or  beneficiaries. As a consequence the traceability of blockchain transactions is  prevented. Hence, shielded addresses are able to perform coin mixing \cite{coinjoin}. Currently, the two largest blockchains (measured by market capitalization) implementing shielded adresses are Zcash \cite{zerocash} and Monero \cite{cryptonote}. While Zcash leverages zero-knowledge succinct non-interactive arguments of knowledge (zk-SNARKs), Monero employs linkable ring signatures (LRS) along with one-time public addresses.

\begin{figure}
	\centering
	\def\svgwidth{\columnwidth}
	\tiny{
		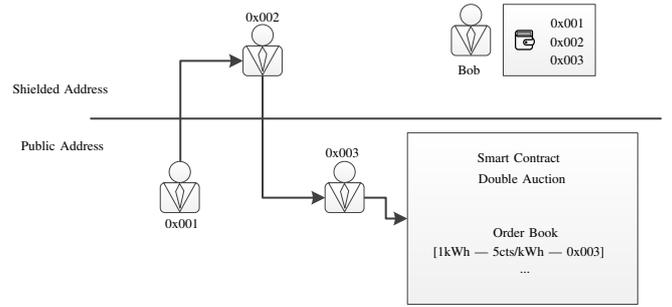
	}	
	\caption{Shielded transactions break linkability among successive orders to preserve user privacy.}
	\label{fig:shieldedmarketplace}
\end{figure}

Fig. \ref{fig:shieldedmarketplace} shows the proposed generic transaction flow. $0x001$ may be a public address, possibly funded by an exchange and therefore linkable to an identity. The user \emph{Bob} sends funds to a shielded address $0x002$, thereby breaking the link to his identity. For each order, Bob sends the necessary value to a fresh unshielded address $0x003$, which in turn is used to call the smart contract to register his order. 

Both Zcash and Monero use unspent-transaction-outputs (UTXO) \cite{nakamoto08} but they don't feature smart contracts. Ethereum \cite{ethereum} on the other hand does feature smart contracts but uses account-based transactions. Therefore, it is known who called a smart contract as only the payload of the contract call can be encrypted.

Quorum \cite{quorum} built a hybrid blockchain featuring a modified Zerocoin protocol along with smart contracts.
While our proposed concept is expected to work on Quorum, zk-SNARKS are computationally heavy and applying such a protocol to embedded applications like our blockchain-smart-meter appears unreasonable today. However, recent advancements in the field of cryptographic primitives led to the release of JubJub \cite{jubjub} enabling faster zero-knowledge proof generation with less RAM usage. 

The CryptoNote protocol \cite{cryptonote} is considered to be more suitable for embedded applications because no heavy computation is involved. If implemented on top of Tendermint, shielded transactions can be combined with smart contracts.

Both Zerocoin and CryptoNote risk leaking private information when using light clients. For both protocols, every user has to scan every block for transactions concerning himself. If a light client queries a full node for his account balance, the full node learns the user's individual public keys. For \Quartierstrom this is a drawback, as pure consumers who do not sell excess energy are equipped with cheaper light client devices.

%% file: privacy_preserving_double_auction_sketch_NME_20190227.pdf_tex
\begingroup%
  \makeatletter%
  \providecommand\color[2][]{%
    \errmessage{(Inkscape) Color is used for the text in Inkscape, but the package 'color.sty' is not loaded}%
    \renewcommand\color[2][]{}%
  }%
  \providecommand\transparent[1]{%
    \errmessage{(Inkscape) Transparency is used (non-zero) for the text in Inkscape, but the package 'transparent.sty' is not loaded}%
    \renewcommand\transparent[1]{}%
  }%
  \providecommand\rotatebox[2]{#2}%
  \newcommand*\fsize{\dimexpr\f@size pt\relax}%
  \newcommand*\lineheight[1]{\fontsize{\fsize}{#1\fsize}\selectfont}%
  \ifx\svgwidth\undefined%
    \setlength{\unitlength}{332.86303627bp}%
    \ifx\svgscale\undefined%
      \relax%
    \else%
      \setlength{\unitlength}{\unitlength * \real{\svgscale}}%
    \fi%
  \else%
    \setlength{\unitlength}{\svgwidth}%
  \fi%
  \global\let\svgwidth\undefined%
  \global\let\svgscale\undefined%
  \makeatother%
  \begin{picture}(1,0.45294647)%
    \lineheight{1}%
    \setlength\tabcolsep{0pt}%
    \put(0,0){\includegraphics[width=\unitlength,page=1]{privacy_preserving_double_auction_sketch_NME_20190227.pdf}}%
    \put(0.03989623,0.31507003){\color[rgb]{0,0,0}\makebox(0,0)[lt]{\lineheight{1.25}\smash{\begin{tabular}[t]{l}Shielded Address\end{tabular}}}}%
    \put(0,0){\includegraphics[width=\unitlength,page=2]{privacy_preserving_double_auction_sketch_NME_20190227.pdf}}%
    \put(0.05239384,0.229909){\color[rgb]{0,0,0}\makebox(0,0)[lt]{\lineheight{1.25}\smash{\begin{tabular}[t]{l}Public Address\end{tabular}}}}%
    \put(0,0){\includegraphics[width=\unitlength,page=3]{privacy_preserving_double_auction_sketch_NME_20190227.pdf}}%
    \put(0.69851993,0.34359231){\color[rgb]{0,0,0}\makebox(0,0)[lt]{\lineheight{1.25}\smash{\begin{tabular}[t]{l}Bob\end{tabular}}}}%
    \put(0,0){\includegraphics[width=\unitlength,page=4]{privacy_preserving_double_auction_sketch_NME_20190227.pdf}}%
    \put(0.8359011,0.41259849){\color[rgb]{0,0,0}\makebox(0,0)[lt]{\lineheight{1.25}\smash{\begin{tabular}[t]{l}0x001\end{tabular}}}}%
    \put(0.83537532,0.38519018){\color[rgb]{0,0,0}\makebox(0,0)[lt]{\lineheight{1.25}\smash{\begin{tabular}[t]{l}0x002\end{tabular}}}}%
    \put(0.83550522,0.35778761){\color[rgb]{0,0,0}\makebox(0,0)[lt]{\lineheight{1.25}\smash{\begin{tabular}[t]{l}0x003\\\end{tabular}}}}%
    \put(0,0){\includegraphics[width=\unitlength,page=5]{privacy_preserving_double_auction_sketch_NME_20190227.pdf}}%
    \put(0.5027808,0.2200208){\color[rgb]{0,0,0}\makebox(0,0)[lt]{\lineheight{1.25}\smash{\begin{tabular}[t]{l}0x003\end{tabular}}}}%
    \put(0,0){\includegraphics[width=\unitlength,page=6]{privacy_preserving_double_auction_sketch_NME_20190227.pdf}}%
    \put(0.38483667,0.42170255){\color[rgb]{0,0,0}\makebox(0,0)[lt]{\lineheight{1.25}\smash{\begin{tabular}[t]{l}0x002\end{tabular}}}}%
    \put(0,0){\includegraphics[width=\unitlength,page=7]{privacy_preserving_double_auction_sketch_NME_20190227.pdf}}%
    \put(0.26537366,0.11442873){\color[rgb]{0,0,0}\makebox(0,0)[lt]{\lineheight{1.25}\smash{\begin{tabular}[t]{l}0x001\end{tabular}}}}%
    \put(0,0){\includegraphics[width=\unitlength,page=8]{privacy_preserving_double_auction_sketch_NME_20190227.pdf}}%
    \put(0.72731465,0.21156572){\color[rgb]{0,0,0}\makebox(0,0)[lt]{\lineheight{1.25}\smash{\begin{tabular}[t]{l}Smart Contract\end{tabular}}}}%
    \put(0.72879236,0.18134175){\color[rgb]{0,0,0}\makebox(0,0)[lt]{\lineheight{1.25}\smash{\begin{tabular}[t]{l}Double Auction\end{tabular}}}}%
    \put(0.75075614,0.10067577){\color[rgb]{0,0,0}\makebox(0,0)[lt]{\lineheight{1.25}\smash{\begin{tabular}[t]{l}Order Book\end{tabular}}}}%
    \put(0.6617701,0.0727665){\color[rgb]{0,0,0}\makebox(0,0)[lt]{\lineheight{1.25}\smash{\begin{tabular}[t]{l}[1kWh | 5cts/kWh | 0x003]\end{tabular}}}}%
    \put(0.79053281,0.04784734){\color[rgb]{0,0,0}\makebox(0,0)[lt]{\lineheight{1.25}\smash{\begin{tabular}[t]{l}...\end{tabular}}}}%
  \end{picture}%
\endgroup%

%% file: tee-auction.tex
\begin{figure}
	\centering
	\def\svgwidth{\columnwidth}
	\tiny{
		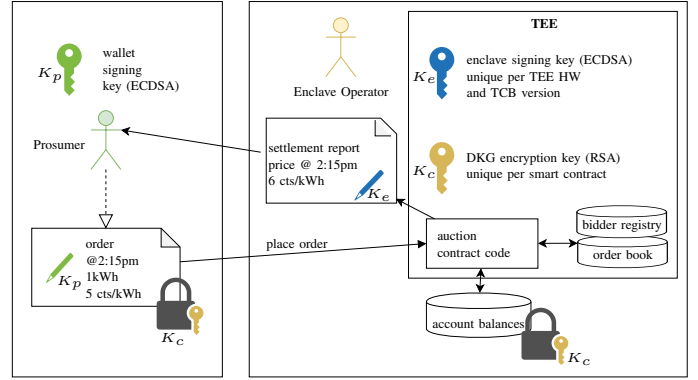
	}	
	\caption{Basic auction setup with a single trusted execution environment acting as trusted auctioneer.}
	\label{fig:teeauction}
\end{figure}

Trusted execution environments (TEEs) allow confidential code execution inside an enclave that is secured by hardware. Such TEEs convince third parties of the integrity and confidentiality of a computation by means of remote attestation by the hardware manufacturer. \cite{sawtoothpdo}, \cite{ekiden}, \cite{lind17} and \cite{encointer} proposed protocols for private blockchain transactions based on Intel SGX \cite{intelsgxexplained}. For embedded devices, ARM TrustZone \cite{trustzone} could be leveraged. Keystone \cite{keystone} develops an open source RISC V CPU featuring a TEE which, unlike the former two, will be formally verifiable. In any of these cases one has to trust the device manufacturer.

Fig. \ref{fig:teeauction} shows the proposed architecture for a confidential double auction, where confidentiality is ensured by TEEs. A prosumer's smart meter creates an order based on the energy measurement for the last 15-min-slot and the customer's price preference. It signs the order with a wallet key $K_p$, allowing the auction enclave to access his pre-paid funds. The prosumer encrypts his order with an encryption key $K_c$ provided by the auction enclave. Therefore, no one except the auction enclave can decrypt the order and learn about its content or originator.

The auction enclave collects all orders in its order book and clears the market. The auction enclave maintains a ledger with all pre-paid account balances and settles the market immediately.
A public settlement report, signed by the enclave key $K_e$, is finally broadcast to all bidders to provide price transparency.

So far, this does not involve a blockchain at all. The described concept could be applied using a central enclave operator. The central operator would not be able to read any orders or account balances. It could even manage account refills supporting standard payment options. To gain a customer's trust, the operator should allow remote attestation by the TEE manufacturer and the auction enclave code would have to be open sourced and would have to be built deterministically so anyone can build the trusted computing base (TCB) and compare its hash to the remote attestation's quote. 

This centralized \emph{trusted computing} approach already comes with privacy benefits if compared to today's standard meter-to-cash process, because confidentiality and integrity of the market are enforced by hardware. However, \Quartierstrom aims to provide a \emph{decentralized} market for local communities. Decentralizing the described market involves allowing anyone to act as enclave operator and the settlement layer may not rely on a trusted third party. 

\begin{figure}
	\centering
	\def\svgwidth{\columnwidth}
	\small{
		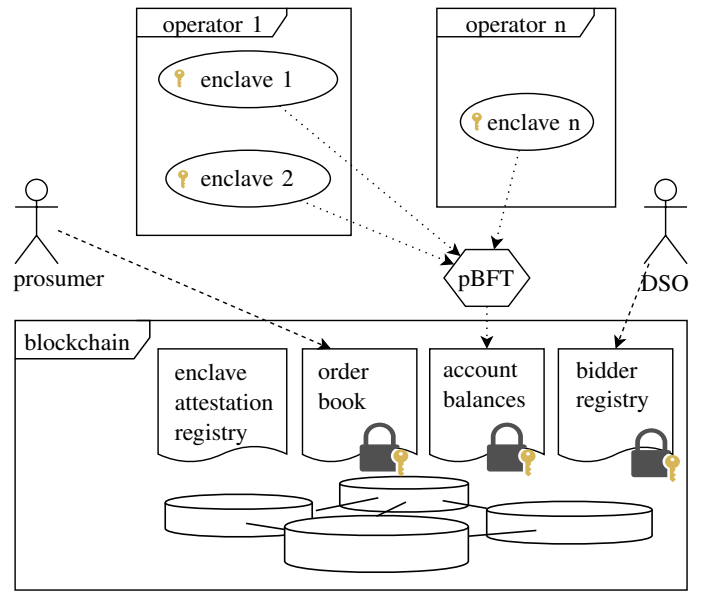
	}	
	\caption{Blockchain setup with off-chain TEEs who register with a registry contract. A DSO registers prosumers allowed to trade. Prosumers send their orders to the blockchain in encrypted form. The enclaves may perform token transactions on behalf of prosumers to settle the market.}
	\label{fig:offchaintee}
\end{figure}

Fig. \ref{fig:offchaintee} shows a setup leveraging blockchain for a decentralized registry of attested enclaves: (i) A TEE operator registers an auction enclave by supplying a remote attestation (RA) quote to the enclave attestation registry (ER) (on-chain). (ii) The ER validates the RA quote and updates its registry. (iii) All registered enclaves generate a new distributed key pair $K_c$ (see \ref{sec:TAconf}). 

\subsection{Trusted Auctioneer Integrity}
Integrity of computation is guaranteed by the TEE manufacturer. However,  successful side-channel attacks have been shown in \cite{foreshadow} compromising not only confidentiality but also integrity. To mitigate such attacks, we suggest to diversify by running a pBFT consensus \cite{castro1999} among multiple TEEs from different manufacturers. \cite{encointer} proposes an incentive scheme to balance power among different manufacturers. 

\subsection{Trusted Auctioneer Confidentiality}\label{sec:TAconf}
If more than one TEE may act as auctioneer, a shared secret needs to be established, known to all registered enclaves but no one else. \cite{ekiden} suggests to use a distributed key generation (DKG) protocol published in \cite{gennaro1999}. Such a shared secret can then be used to derive the encryption key pair $K_c$ in Fig. \ref{fig:teeauction}.
Frequent renewal of the shared secret is recommended to improve forward-secrecy in the case of future TEE vulnerabilities. 

\subsection{Settlement}
An auction enclave must be able to settle the market without any interaction with the prosumer. It therefore is a custodian of a prepaid balance on behalf of the latter. In our  decentralized scenario there is no single operator that could offer a prepaid account topup service to the prosumers as in the case of a centralized auctioneer.   
A practical approach would be that utilities are permissioned to issue IOUs ("I owe you") denominated in fiat currency on behalf of prosumers topping up their balance. 
A more decentralized way would be to introduce bridges to cryptocurrency blockchains along the ideas presented in \cite{tesseract}. 

\subsection{Permissioned Bidder Registry}
There is no way around the natural monopoly of a distribution system operator (DSO). Only the DSO knows which prosumers are connected to the same low-voltage grid and should therefore be able to trade energy on their local market as envisioned by \Quartierstrom. The DSO also knows who actually operates a real PV plant and of what size. Therefore, decentralization faces its natural limits when touching physical givens, represented by the bidder registry.
 


%% file: double-auction-off-chain-TEE-inkscape.pdf_tex
\begingroup%
  \makeatletter%
  \providecommand\color[2][]{%
    \errmessage{(Inkscape) Color is used for the text in Inkscape, but the package 'color.sty' is not loaded}%
    \renewcommand\color[2][]{}%
  }%
  \providecommand\transparent[1]{%
    \errmessage{(Inkscape) Transparency is used (non-zero) for the text in Inkscape, but the package 'transparent.sty' is not loaded}%
    \renewcommand\transparent[1]{}%
  }%
  \providecommand\rotatebox[2]{#2}%
  \newcommand*\fsize{\dimexpr\f@size pt\relax}%
  \newcommand*\lineheight[1]{\fontsize{\fsize}{#1\fsize}\selectfont}%
  \ifx\svgwidth\undefined%
    \setlength{\unitlength}{509.25000288bp}%
    \ifx\svgscale\undefined%
      \relax%
    \else%
      \setlength{\unitlength}{\unitlength * \real{\svgscale}}%
    \fi%
  \else%
    \setlength{\unitlength}{\svgwidth}%
  \fi%
  \global\let\svgwidth\undefined%
  \global\let\svgscale\undefined%
  \makeatother%
  \begin{picture}(1,0.5611193)%
    \lineheight{1}%
    \setlength\tabcolsep{0pt}%
    \put(0,0){\includegraphics[width=\unitlength,page=1]{double-auction-off-chain-TEE-inkscape.pdf}}%
    \put(0.48748158,0.41973491){\color[rgb]{0,0,0}\makebox(0,0)[t]{\lineheight{1.25}\smash{\begin{tabular}[t]{c}Enclave Operator\end{tabular}}}}%
    \put(0,0){\includegraphics[width=\unitlength,page=2]{double-auction-off-chain-TEE-inkscape.pdf}}%
    \put(0.7879234,0.52135494){\makebox(0,0)[t]{\lineheight{1.25}\smash{\begin{tabular}[t]{c}\textbf{TEE}\end{tabular}}}}%
    \put(0,0){\includegraphics[width=\unitlength,page=3]{double-auction-off-chain-TEE-inkscape.pdf}}%
    \put(0.42204922,0.19314115){\color[rgb]{0,0,0}\makebox(0,0)[t]{\lineheight{1.25}\smash{\begin{tabular}[t]{c}place order\end{tabular}}}}%
    \put(0,0){\includegraphics[width=\unitlength,page=4]{double-auction-off-chain-TEE-inkscape.pdf}}%
    \put(0.90427097,0.1759418){\color[rgb]{0,0,0}\makebox(0,0)[t]{\lineheight{1.25}\smash{\begin{tabular}[t]{c}order book\end{tabular}}}}%
    \put(0,0){\includegraphics[width=\unitlength,page=5]{double-auction-off-chain-TEE-inkscape.pdf}}%
    \put(0.1346518,0.47759312){\color[rgb]{0,0,0}\makebox(0,0)[lt]{\lineheight{1.25}\smash{\begin{tabular}[t]{l}wallet \\signing \\key (ECDSA)\end{tabular}}}}%
    \put(0,0){\includegraphics[width=\unitlength,page=6]{double-auction-off-chain-TEE-inkscape.pdf}}%
    \put(0.07006102,0.34125816){\color[rgb]{0,0,0}\makebox(0,0)[t]{\lineheight{1.25}\smash{\begin{tabular}[t]{c}Prosumer\end{tabular}}}}%
    \put(0,0){\includegraphics[width=\unitlength,page=7]{double-auction-off-chain-TEE-inkscape.pdf}}%
    \put(0.69072165,0.07363772){\color[rgb]{0,0,0}\makebox(0,0)[t]{\lineheight{1.25}\smash{\begin{tabular}[t]{c}account balances\end{tabular}}}}%
    \put(0,0){\includegraphics[width=\unitlength,page=8]{double-auction-off-chain-TEE-inkscape.pdf}}%
    \put(0.1094046,0.19250998){\color[rgb]{0,0,0}\makebox(0,0)[lt]{\lineheight{1.25}\smash{\begin{tabular}[t]{l}order\\@2:15pm\\1kWh\\5 cts/kWh\end{tabular}}}}%
    \put(0.37976015,0.33557749){\color[rgb]{0,0,0}\makebox(0,0)[lt]{\lineheight{1.25}\smash{\begin{tabular}[t]{l}settlement report\\price @ 2:15pm\\6 cts/kWh\end{tabular}}}}%
    \put(0.62958342,0.20481802){\color[rgb]{0,0,0}\makebox(0,0)[lt]{\lineheight{1.25}\smash{\begin{tabular}[t]{l}auction\\contract code\end{tabular}}}}%
    \put(0.67325898,0.46707348){\color[rgb]{0,0,0}\makebox(0,0)[lt]{\lineheight{1.25}\smash{\begin{tabular}[t]{l}enclave signing key (ECDSA)\\unique per TEE HW \\and TCB version\end{tabular}}}}%
    \put(0.67273255,0.32190199){\color[rgb]{0,0,0}\makebox(0,0)[lt]{\lineheight{1.25}\smash{\begin{tabular}[t]{l}DKG encryption key (RSA)\\unique per smart contract\end{tabular}}}}%
    \put(0.03719265,0.44779959){\color[rgb]{0,0,0}\makebox(0,0)[lt]{\lineheight{1.25}\smash{\begin{tabular}[t]{l}$K_p$\end{tabular}}}}%
    \put(0.06645656,0.13846196){\color[rgb]{0,0,0}\makebox(0,0)[lt]{\lineheight{1.25}\smash{\begin{tabular}[t]{l}$K_p$\end{tabular}}}}%
    \put(0.59171752,0.44263413){\color[rgb]{0,0,0}\makebox(0,0)[lt]{\lineheight{1.25}\smash{\begin{tabular}[t]{l}$K_e$\end{tabular}}}}%
    \put(0.52398673,0.2671485){\color[rgb]{0,0,0}\makebox(0,0)[lt]{\lineheight{1.25}\smash{\begin{tabular}[t]{l}$K_e$\end{tabular}}}}%
    \put(0.59171752,0.30136576){\color[rgb]{0,0,0}\makebox(0,0)[lt]{\lineheight{1.25}\smash{\begin{tabular}[t]{l}$K_c$\end{tabular}}}}%
    \put(0.82069599,0.02316469){\color[rgb]{0,0,0}\makebox(0,0)[lt]{\lineheight{1.25}\smash{\begin{tabular}[t]{l}$K_c$\end{tabular}}}}%
    \put(0.2196627,0.052175){\color[rgb]{0,0,0}\makebox(0,0)[lt]{\lineheight{1.25}\smash{\begin{tabular}[t]{l}$K_c$\end{tabular}}}}%
    \put(0,0){\includegraphics[width=\unitlength,page=9]{double-auction-off-chain-TEE-inkscape.pdf}}%
    \put(0.90303018,0.22061921){\color[rgb]{0,0,0}\makebox(0,0)[t]{\lineheight{1.25}\smash{\begin{tabular}[t]{c}bidder registry\end{tabular}}}}%
  \end{picture}%
\endgroup%

%% file: blockchain-registry-tee-inkscape.pdf_tex
\begingroup%
  \makeatletter%
  \providecommand\color[2][]{%
    \errmessage{(Inkscape) Color is used for the text in Inkscape, but the package 'color.sty' is not loaded}%
    \renewcommand\color[2][]{}%
  }%
  \providecommand\transparent[1]{%
    \errmessage{(Inkscape) Transparency is used (non-zero) for the text in Inkscape, but the package 'transparent.sty' is not loaded}%
    \renewcommand\transparent[1]{}%
  }%
  \providecommand\rotatebox[2]{#2}%
  \newcommand*\fsize{\dimexpr\f@size pt\relax}%
  \newcommand*\lineheight[1]{\fontsize{\fsize}{#1\fsize}\selectfont}%
  \ifx\svgwidth\undefined%
    \setlength{\unitlength}{354.60534668bp}%
    \ifx\svgscale\undefined%
      \relax%
    \else%
      \setlength{\unitlength}{\unitlength * \real{\svgscale}}%
    \fi%
  \else%
    \setlength{\unitlength}{\svgwidth}%
  \fi%
  \global\let\svgwidth\undefined%
  \global\let\svgscale\undefined%
  \makeatother%
  \begin{picture}(1,0.86927623)%
    \lineheight{1}%
    \setlength\tabcolsep{0pt}%
    \put(0,0){\includegraphics[width=\unitlength,page=1]{blockchain-registry-tee-inkscape.pdf}}%
    \put(0.09688334,0.35955465){\color[rgb]{0,0,0}\makebox(0,0)[t]{\lineheight{1.25}\smash{\begin{tabular}[t]{c}blockchain\end{tabular}}}}%
    \put(0,0){\includegraphics[width=\unitlength,page=2]{blockchain-registry-tee-inkscape.pdf}}%
    \put(0.29781093,0.83332077){\color[rgb]{0,0,0}\makebox(0,0)[t]{\lineheight{1.25}\smash{\begin{tabular}[t]{c}operator 1\end{tabular}}}}%
    \put(0,0){\includegraphics[width=\unitlength,page=3]{blockchain-registry-tee-inkscape.pdf}}%
    \put(0.74619672,0.83332077){\color[rgb]{0,0,0}\makebox(0,0)[t]{\lineheight{1.25}\smash{\begin{tabular}[t]{c}operator n\end{tabular}}}}%
    \put(0,0){\includegraphics[width=\unitlength,page=4]{blockchain-registry-tee-inkscape.pdf}}%
    \put(0.70178115,0.45473088){\color[rgb]{0,0,0}\makebox(0,0)[t]{\lineheight{1.25}\smash{\begin{tabular}[t]{c}pBFT\end{tabular}}}}%
    \put(0,0){\includegraphics[width=\unitlength,page=5]{blockchain-registry-tee-inkscape.pdf}}%
    \put(0.0025013,0.45408438){\color[rgb]{0,0,0}\makebox(0,0)[lt]{\lineheight{1.25}\smash{\begin{tabular}[t]{l}prosumer\end{tabular}}}}%
    \put(0,0){\includegraphics[width=\unitlength,page=6]{blockchain-registry-tee-inkscape.pdf}}%
    \put(0.96615957,0.44627077){\color[rgb]{0,0,0}\makebox(0,0)[t]{\lineheight{1.25}\smash{\begin{tabular}[t]{c}DSO\end{tabular}}}}%
    \put(0,0){\includegraphics[width=\unitlength,page=7]{blockchain-registry-tee-inkscape.pdf}}%
    \put(0.77272234,0.68738388){\color[rgb]{0,0,0}\makebox(0,0)[t]{\lineheight{1.25}\smash{\begin{tabular}[t]{c}enclave n\end{tabular}}}}%
    \put(0,0){\includegraphics[width=\unitlength,page=8]{blockchain-registry-tee-inkscape.pdf}}%
    \put(0.34857158,0.7508347){\color[rgb]{0,0,0}\makebox(0,0)[t]{\lineheight{1.25}\smash{\begin{tabular}[t]{c}enclave 1\end{tabular}}}}%
    \put(0,0){\includegraphics[width=\unitlength,page=9]{blockchain-registry-tee-inkscape.pdf}}%
    \put(0.34857158,0.60278279){\color[rgb]{0,0,0}\makebox(0,0)[t]{\lineheight{1.25}\smash{\begin{tabular}[t]{c}enclave 2\end{tabular}}}}%
    \put(0,0){\includegraphics[width=\unitlength,page=10]{blockchain-registry-tee-inkscape.pdf}}%
    \put(0.2409198,0.31384687){\color[rgb]{0,0,0}\makebox(0,0)[lt]{\lineheight{1.25}\smash{\begin{tabular}[t]{l}enclave\\attestation\\registry \end{tabular}}}}%
    \put(0.63811197,0.31905602){\color[rgb]{0,0,0}\makebox(0,0)[lt]{\lineheight{1.25}\smash{\begin{tabular}[t]{l}account \\balances\end{tabular}}}}%
    \put(0.83475465,0.31775368){\color[rgb]{0,0,0}\makebox(0,0)[lt]{\lineheight{1.25}\smash{\begin{tabular}[t]{l}bidder \\registry\end{tabular}}}}%
    \put(0.45318972,0.31514921){\color[rgb]{0,0,0}\makebox(0,0)[lt]{\lineheight{1.25}\smash{\begin{tabular}[t]{l}order \\book\end{tabular}}}}%
  \end{picture}%
\endgroup%

%% file: related-work.tex
The methods presented in Section \ref{sec:shieldedauction} extend and concretize the ideas of \cite{laska2018}. While \cite{laska2018} performs computations off-chain and uses the blockchain as a log, our approach uses completely on-chain computations for the shielded auction.

Ekiden \cite{ekiden} is a platform for confidentiality-preserving, trustworthy and performant smart contracts built on TEEs. It is a universal platform built on Tendermint and a possible option for an implementation of a prototype. 


Strain (Secure aucTions foR blockchAINs) \cite{strain} proposes a sealed-bid auction for blockchains, leveraging zero-knowledge proofs. The protocol, however, relies on a semi-trusted judge.